\documentclass[aps,prd,twocolumn]{revtex4}

\begin{document}

\title{Symmetric hyperbolic form of systems of second-order evolution
equations subject to constraints}

\author{Carsten Gundlach}
\email[]{C.Gundlach@maths.soton.ac.uk} 
%\homepage[]{http://www.maths.soton.ac.uk/staff/Gundlach}
\affiliation{School of Mathematics, University of Southampton,
         Southampton SO17 1BJ, UK}

\author{Jos\'e M. Mart\'\i n-Garc\'\i a}
\email[]{jmm@imaff.cfmac.csic.es} 
%\homepage[]{http://metric.imaff.csic.es/Martin-Garcia}
\affiliation{\mbox{Instituto de Matem\'aticas y F\'{\i}sica Fundamental,
Centro de F\'{\i}sica Miguel A. Catal\'an,} \\
C.S.I.C., Serrano 113 bis, 28006 Madrid, Spain}

\date{18 February 2004}

%%%%%%%%%%%%%%%%%%%%%%%%%%%%%%%%%%%%%%%%%%%%%%%%%%%%%%%%%%%%%%%%%%%%%%%%

\begin{abstract}

Motivated by the initial-boundary value problem for the Einstein
equations, we propose a definition of symmetric hyperbolicity for
systems of evolution equations that are first order in time but second
order in space. This can be used to impose constraint-preserving
boundary conditions. The general methods are illustrated in detail in
the toy model of electromagnetism.

\end{abstract}

%%%%%%%%%%%%%%%%%%%%%%%%%%%%%%%%%%%%%%%%%%%%%%%%%%%%%%%%%%%%%%%%%%%%%%%%

\maketitle

%\tableofcontents

%%%%%%%%%%%%%%%%%%%%%%%%%%%%%%%%%%%%%%%%%%%%%%%%%%%%%%%%%%%%%%%%%%%%%%%%

\section{Introduction}

%%%%%%%%%%%%%%%%%%%%%%%%%%%%%%%%%%%%%%%%%%%%%%%%%%%%%%%%%%%%%%%%%%%%%%%%

In attempting to solve the Einstein equations numerically as an
initial or initial-boundary value problem, it is important to start
with a well-posed continuum problem. For systems of first-order
equations, there are two useful criteria for well-posedness: Strong
hyperbolicity is a necessary and sufficient criterion for the initial
value problem (without boundaries or with periodic boundaries) to be
well-posed. Roughly speaking, it means that the system possesses a
complete set of characteristic variables, which propagate with
definite speeds. Symmetric hyperbolicity implies strong hyperbolicity,
and can be used to formulate a well-posed initial-boundary value
problem.  Roughly speaking, it means that the principal part of the
system possesses an (unphysical) energy whose growth due to boundary
terms can be expressed as ``incoming mode energy minus outgoing mode
energy''.

Much less is known about how to make systems of second-order equations
well-posed. Here we concentrate on systems which have been brought
into first order in time but remain second order in space. The
simplest example of such a system is the wave equation, written as
\begin{eqnarray}
\label{wave1}
\partial_t\phi&=&\Pi, \\
\label{wave2}
\partial_t\Pi&=&\partial_i\partial^i\phi.
\end{eqnarray}
The Arnowitt-Deser-Misner (ADM) formulation of the Einstein equations has a
similar form, with the 3-metric corresponding to $\phi$ and the
extrinsic curvature to $\Pi$. 

In this paper we discuss three methods for investigating and
proving well-posedness for such systems: we review two known methods in
order to put our work into context, and then propose one which we
believe to be new and advantageous. 

The first old method, (differential) reduction to first order,
introduces the spatial derivatives as auxiliary variables. For the
wave equation above, these could be called $d_i\equiv \partial_i
\phi$. One then has a larger but first-order system, which one can
hope to make strongly or symmetric hyperbolic. A drawback of this
method is that its solution space is larger than that of the original
system: it is equivalent to the original system only if the three
auxiliary constraints $C_i\equiv d_i-\partial_i \phi$
vanish. Therefore proving strong or symmetric hyperbolicity for the
first-order reduction does not prove well-posedness for the original
second-order system except in simple cases where the auxiliary
constraints evolve trivially. (These include the wave equation and
electromagnetism, but not the full Einstein equations.)

A second method called pseudo-differential reduction to first order
Fourier-transforms the PDE system in space. If it is linear with
constant coefficients, one effectively obtains a system of decoupled
ODEs, one for each wave number $\omega_i$. One can then show a version
of strong hyperbolicity for this system. The key point is that the
pseudo-differential reduction does not introduce any auxiliary
variables. Proving strong hyperbolicity therefore also proves
well-posedness of the initial value problem for the original
second-order system \cite{KreissOrtiz}. This method has been applied
to a formulation of the Einstein equations in \cite{NOR}. A drawback
is that as the Fourier transform is nonlocal, one cannot find a
locally conserved energy in this way, and so cannot address the
initial-boundary value problem.

In order to avoid the drawbacks of these two approaches, we propose a
fairly obvious generalisation of the concept of symmetric
hyperbolicity to systems that are second order in space and first
order in time. In the example of the wave equation, we define
characteristic variables in a given direction $n_i$ as $\Pi\pm
n^i\partial_i\phi$, and an energy density
$\Pi^2+(\partial_i\phi)^2$. The energy and its associated flux can be
written in terms of characteristic variables, and so we can impose
maximally dissipative boundary conditions similar to those for a
first-order symmetric hyperbolic system.

The paper is organised as follows: In Sec.~\ref{section:firstorder} we
review strong and symmetric hyperbolicity, and the initial-boundary
value problem, for first-order systems. In Sec.~\ref{section:pseudo}
we review pseudo-differential reduction to first order. In
Sec.~\ref{section:secondorder} we present our concept of symmetric
hyperbolicity for a system that is second order in space and first
order in time. In Sec.~\ref{section:cpbc} we review the
initial-boundary value problem for systems with differential
constraints, such as the Maxwell or Einstein equations. This
discussion applies both to first and to second order symmetric
hyperbolic systems. 

Finally, in Sec.~\ref{section:maxwell} we compare the three different
methods by applying them to the Maxwell equations: reduction to first
order with auxiliary variables, pseudo-differential reduction to first
order, and our method of finding characteristic variables and an
energy directly for the second-order system. We formulate
constraint-preserving boundary conditions, show that the
initial-boundary value problem for the second-order Maxwell equations
can be made well-posed for ``Neumann'' and ``Dirichlet'' boundary
conditions, and conjecture that it is well-posed for a one-parameter
family of boundary conditions which interpolates between those two
via ``Sommerfeld'' boundary conditions.

%%%%%%%%%%%%%%%%%%%%%%%%%%%%%%%%%%%%%%%%%%%%%%%%%%%%%%%%%%%%%%%%%%%%%%%%

\section{First-order systems}
\label{section:firstorder}

In this section we summarise some general results on the
well-posedness of systems of first-order evolution equations that are
proved for example in \cite{GustafssonKreiss}. We begin by considering
linear systems with constant coefficients, and at the end of the
section appeal to standard methods for generalising these results to
quasilinear systems. (For other reviews, see \cite{Stewart,Reula}.)

%%%%%%%%%%%%%%%%%%%%%%%%%%%%%%%%%%%%%%%%%%%%%%%%%%%%%%%%%%%%%%%%%%%%%%%%

\subsection{Strong hyperbolicity}

%%%%%%%%%%%%%%%%%%%%%%%%%%%%%%%%%%%%%%%%%%%%%%%%%%%%%%%%%%%%%%%%%%%%%%%%

Consider the system of first-order linear evolution equations with
constant coefficients
\begin{equation}
\partial_t u = P^i \partial_i u + Qu.
\end{equation}
Here $u$ is a vector of variables, $P^i$ and $Q$ are square
matrices, and $i$ ranges over the spatial directions. The
initial-value problem for such a system is
called well-posed if a solution exists, is unique and depends
continuously on the initial data $u(x,0)$. This is equivalent to there
being an estimate
\begin{equation}
\label{estimates}
||u(\cdot,t)|| \le Ke^{\alpha t} ||u(\cdot,0)||
\end{equation}
with respect to an $L_2$ norm, where the constants $K$ and $\alpha$ do
not depend on the initial data. To simplify the presentation, in the
following we shall assume that $Q=0$. Leaving $Q$ in the calculation
would show that it influences the value of $\alpha$, but not the
existence of the estimate itself.

The main result is that the existence of an estimate (\ref{estimates})
is equivalent to the system being {\it strongly hyperbolic}: for any
unit covector $n_i$, the matrix $P_n= n_iP^i$ has only real
eigenvalues and a complete set of real eigenvectors.

The idea of the proof of this theorem is to go to Fourier space. Let
$\hat u(\omega,t)$ be the Fourier transform of $u(x,t)$:
\begin{equation}
u(x,t)\equiv  \int_{R^3} \hat u(\omega,t) e^{i\omega_i
x^i}{d^3\omega\over (2\pi)^{3/2}}.  
\end{equation}
Parseval's theorem states that $||u||_x=||\hat u||_\omega$ and so the
estimate (\ref{estimates}) is equivalent to
\begin{equation}
\label{Fourierestimate}
||\hat u(\cdot,t)|| \le Ke^{\alpha t} ||\hat u(\cdot,0)||
\end{equation}
The system in $x$ and $t$
becomes an ODE system for each value of $\omega_i$:
\begin{equation}
\label{pseudodifferentialODE}
\partial_t \hat u = i|\omega| P_n \hat u,
\end{equation}
where now $n_i=\omega_i/|\omega|$. Now, if $P_n$ has complex eigenvalues,
then $\alpha$ depends on the largest $|\omega|$ present in the initial
data, and hence on the initial data. If $P_n$ lacks a complete set of
eigenvectors, the normal form of $P_n$ contains at least one Jordan
block of size $k>1$. One can show that then $|\hat u(\omega,t)|$
is bounded only by $|\omega t|^{k-1}$, so that now $K$ depends on the
largest value of $|\omega|$ present in the initial data.

On the other hand, {\it if} $P_n$ is diagonalisable with real
eigenvalues, let $\Lambda_n$ be the diagonal matrix of eigenvalues, and
$T_n$ be a matrix of the corresponding right (column) eigenvectors of
$P_n$,
\begin{equation}
P_n T_n = T_n \Lambda_n. 
\end{equation}
(Note that $T_n$ is not unique.) 
The Fourier characteristic variables
\begin{equation}
\hat U(\omega,t)\equiv T_n^{-1} \hat u(\omega,t)
\end{equation}
obey 
\begin{equation}
\partial_t \hat U=i|\omega|\Lambda_n \hat U.
\end{equation}
This gives
\begin{equation}
|\hat U(\omega,t)|=|\hat U(\omega,0)|,
\end{equation}
and so
\begin{equation}
|\hat u(\omega,t)|\le |T_n|\,|T_n^{-1}|\,|\hat u(\omega,0)|.
\end{equation}
If $T_n$ and $T_n^{-1}$ are continuous in $n_i$, we obtain
(\ref{Fourierestimate}) and hence (\ref{estimates}), with
$\alpha=0$ and $K=\sup( |T_n|\,|T_n^{-1}|)$.

We now go back to real space, and define the vector $U$ of
characteristic variables with respect to an arbitrary fixed direction
$n_i$ by
\begin{equation}
U\equiv T_n^{-1} u. 
\end{equation}
The characteristic variables always depend on the direction $n_i$, but
following standard notation we do not indicate this explicitly.  
We define the derivative in the direction $n_i$ as
\begin{equation}
\partial_n\equiv n^i\partial_i,
\end{equation}
and the projector ${q^i}_j$ into the space normal to $n_i$ by
\begin{equation}
{q^i}_j\equiv {\delta^i}_j-n^in_j.
\end{equation}
(Note that we need a spatial metric to define $n^i$.)
We also define the shorthand index notation $A$ for a tensor index $i$
that has been projected with ${q^i}_j$, for example in
\begin{equation}
P^i {q_i}^j \partial_j\equiv P^A \partial_A
\end{equation}
With this notation we can write
\begin{equation}
P^i\partial_i = P_n\partial_n + P^A\partial_A. 
\end{equation}
The characteristic variables obey 
\begin{equation}
\label{Ueqn}
\partial_t U=\Lambda_n \partial_n U + (T_n^{-1}P^A T_n) \partial_A U.
\end{equation}

Strong hyperbolicity implies the existence of a {\it symmetriser}, which
is a Hermitian matrix $H_n$ for every unit covector $n_i$ that obeys
\begin{equation}
(H_n P_n)^\dagger=H_n P_n
\end{equation}
The general matrix $H_n$ with this property can be constructed as
\begin{equation}
H_n=(T_n^{-1})^\dagger B_n T_n^{-1},
\end{equation}
where $B_n$ is Hermitian, positive definite, and commutes with
$\Lambda_n$. The general matrix $B_n$ with this last property is
block-diagonal, with each block corresponding to one (possibly
multiple) eigenvalue in $\Lambda_n$. For given $T_n$, $H_n$ is
parameterised by $B_n$. Conversely, it can be parameterised by $T_n$
with $B_n=I$. If the field equations do not single out a preferred
direction $n_i$, then $\Lambda_n$ do not depend on $n_i$, and it is
natural to choose $T_n$ so that $B_n$ is also independent of $n_i$.
We then write $B$ and $\Lambda$ instead of $B_n$ and $\Lambda_n$.

%%%%%%%%%%%%%%%%%%%%%%%%%%%%%%%%%%%%%%%%%%%%%%%%%%%%%%%%%%%%%%%%%%%%%%%%

\subsection{Symmetric hyperbolicity}

%%%%%%%%%%%%%%%%%%%%%%%%%%%%%%%%%%%%%%%%%%%%%%%%%%%%%%%%%%%%%%%%%%%%%%%%

We shall now see how well-posedness can be maintained in the presence
of boundaries in space. Without loss of generality we assume that
these boundaries are fixed (at constant values of the coordinates
$x^i$.) If a symmetriser $H_n=H$ exists that is independent of $n_i$,
the system is called {\it symmetric hyperbolic}. In this case the
system admits a positive definite {\it energy}
\begin{equation}
\label{defE}
E= \int_\Omega \epsilon\,dV, \qquad 
\epsilon= u^\dagger H u.
\end{equation}
This means that $E$ is conserved up to boundary terms: Assuming still
that $H$ and $P^i$ are constant, and with $HP^i=(HP^i)^\dagger$, we have
\begin{equation}
\partial_t\epsilon
=\partial_i(u^\dagger H P^i u),
\end{equation}
and so by Gauss' law
\begin{equation}
\label{dEdt}
{dE\over dt}=\int_{\partial\Omega} F^i \,dS_i, \qquad
F^i=u^\dagger H P^i u, 
\end{equation}
where $dS_i$ is the surface element on the boundary
$\partial\Omega$. This can be written as
\begin{equation}
\label{dE_char}
{dE\over dt}=\int_{\partial\Omega} F^n\,dS, \qquad
F^n=U^\dagger B \Lambda U,
\end{equation}
where $n_i$ is the normal to $\partial\Omega$.

The boundary flux can be written schematically as ``$U_+^2 -
U_-^2$'', or more precisely as
\begin{equation}
F^n= U_+^\dagger B_+ \Lambda_+ U_+
+U_-^\dagger B_- \Lambda_- U_-,
\end{equation}
where $U_+$ are modes that come in across the boundary and $U_-$ are
outgoing, and where $B_\pm$ and $\Lambda_\pm$ are the blocks
corresponding to $U_\pm$. Therefore $B_+\Lambda_+$ is positive
definite and $B_-\Lambda_-$ is negative definite (the meaning of the
minus sign in ``$U_+^2 - U_-^2$''). There may also be modes $U_0$
whose propagation speed across the boundary is zero. In this case the
boundary is called characteristic. These modes do not contribute to
$F^n$. It is clear that boundary conditions must be imposed on the
incoming variables $U_+$, and that no boundary conditions can be
imposed on the variables $U_-$ and $U_0$. In the following we assume
that the number of variables that propagate along the boundary is
constant.

Estimates of the type (\ref{estimates}) can be obtained from an energy
$E$ whose growth is suitably bounded. One can achieve this by imposing
a {\it maximally dissipative boundary condition}
\begin{equation}
U_+=LU_-+f, 
\end{equation}
where $f$ is given and the matrix $L$ is sufficiently small. For $f=0$
this condition means that more energy goes out than comes in at every
point on the boundary (not just integrated over the boundary). For
$f\ne 0$ given, the growth of $E$ can be bounded from $f$. The
required smallness condition on the matrix $L$ is that the matrix
\begin{equation}
L^\dagger B_+\Lambda_+L+B_-\Lambda_-
\end{equation}
is negative definite. For maximally dissipative boundary conditions to
result in a well-posed initial-boundary value 
problem, the system has to be symmetric
hyperbolic. There are examples where imposing them on a strongly but
not symmetric hyperbolic system leads to an ill-posed problem
\cite{CalabreseSarbach}. 

%%%%%%%%%%%%%%%%%%%%%%%%%%%%%%%%%%%%%%%%%%%%%%%%%%%%%%%%%%%%%%%%%%%%%%%%

\subsection{Variable coefficient and quasilinear systems}

%%%%%%%%%%%%%%%%%%%%%%%%%%%%%%%%%%%%%%%%%%%%%%%%%%%%%%%%%%%%%%%%%%%%%%%%

The results on linear systems with constant coefficients can be
generalised to linear systems with variable coefficients
\cite{GustafssonKreiss}. It is sufficient that the system is strongly
or symmetric hyperbolic at every point in space and time (frozen
coefficients approximation), and bounds can be obtained that are
uniform over all space, all directions $n_i$, and a finite time
interval. In obtaining an energy for the system with variable
coefficients, the time and space derivatives of the non-constant
coefficients give rise to non-principal terms, but these do not affect
well-posedness.

The results for linear problems with variable coefficients can be
extended to nonlinear problems. In a first step, linearising a
non-linear problem around a background solution gives rise to a linear
problem with variable coefficients. Clearly a necessary condition for
the non-linear problem to be well-posed is that the linearised problem
is well-posed. Furthermore, if the linearised problem is well-posed
for all backgrounds, then it may be possible to prove well-posedness
for the nonlinear problem \cite{GustafssonKreiss}. 

Physically, the frozen coefficients, principal part only approximation
means that we focus on possible instabilities in small high-frequency
perturbations. A problem is ill-posed in practice because it allows
perturbation modes whose growth rate is not bounded -- the higher the
frequency in space, the faster the growth in time. In numerical
simulations such modes show up as instabilities at the grid frequency
which grow faster at higher resolution, thus destroying convergence at
sufficiently late times and high resolutions. Well-posedness does not
rule out instabilities (in the loose sense of rapidly growing
perturbations, possibly violating the constraints) which are generated
by non-principal terms and whose growth is bounded independently of
spatial frequency.

%%%%%%%%%%%%%%%%%%%%%%%%%%%%%%%%%%%%%%%%%%%%%%%%%%%%%%%%%%%%%%%%%%%%%%%%

\section{Pseudo-differential reduction to first order}
\label{section:pseudo}

%%%%%%%%%%%%%%%%%%%%%%%%%%%%%%%%%%%%%%%%%%%%%%%%%%%%%%%%%%%%%%%%%%%%%%%%

Pseudo-differential reduction to first order has recently been
suggested as a technique for proving that certain second-order
evolution systems are strongly hyperbolic \cite{KreissOrtiz}. We
review it here only to see how it relates to the two other ways of
dealing with a second-order system: reduction to a first-order system
by introducing auxiliary variables, and the second-order energy method
we propose in this paper.

A system of first order pseudo-differential equations is one whose
Fourier transform in space is of the form
(\ref{pseudodifferentialODE}) but where $P_n$ is {\it not} given
by the linear expression $P_n=n_iP^i$ for a constant
$P^i$. The definition of strong hyperbolicity of a system of evolution
equations generalises to the pseudo-differential case, and strong
hyperbolicity is again equivalent to well-posedness (in the absence of
boundaries), because the proof using the ODE system
(\ref{pseudodifferentialODE}) refers only to the matrix $P_n$ but not
directly to the vector of matrices $P^i$.

First-order pseudo-differential systems can be derived from a class of
second-order differential systems. The simplest example is the wave
equation, in the form (\ref{wave1},\ref{wave2}). The Fourier transform
of this system is
\begin{eqnarray}
\partial_t\hat\phi&=&\hat\Pi, \\
\partial_t\hat\Pi&=&-|\omega|^2\hat\phi.
\end{eqnarray}
Replacing $\hat\phi$ by $\hat\varphi\equiv i|\omega|\hat\phi$
\cite{NOR}, we obtain the first-order pseudo-differential system
\begin{eqnarray}
\partial_t\hat\varphi&=&i|\omega|\hat\Pi, \\
\partial_t\hat\Pi&=&i|\omega|\hat\varphi.
\end{eqnarray}
Note that this is not the Fourier transform of a differential system
as $P_n$ is independent of $n_i$. This system is strongly hyperbolic,
with Fourier characteristic variables
\begin{equation}
\label{scalarfourchar}
\hat U_\pm = \hat\Pi\pm\hat\varphi.
\end{equation}
which obey 
\begin{equation}
\label{dtUhat}
\partial_t \hat U_\pm = \pm i|\omega| \hat U_\pm.
\end{equation}

%%%%%%%%%%%%%%%%%%%%%%%%%%%%%%%%%%%%%%%%%%%%%%%%%%%%%%%%%%%%%%%%%%%%%%%%

\section{Second order in space, first order in time}
\label{section:secondorder}

%%%%%%%%%%%%%%%%%%%%%%%%%%%%%%%%%%%%%%%%%%%%%%%%%%%%%%%%%%%%%%%%%%%%%%%%

We shall now show how the two key ideas for first-order systems --
strong and symmetric hyperbolicity --
can be adapted to a system that is first order in time but second
order in space.  

%%%%%%%%%%%%%%%%%%%%%%%%%%%%%%%%%%%%%%%%%%%%%%%%%%%%%%%%%%%%%%%%%%%%%%%%

\subsection{Strong hyperbolicity}

%%%%%%%%%%%%%%%%%%%%%%%%%%%%%%%%%%%%%%%%%%%%%%%%%%%%%%%%%%%%%%%%%%%%%%%%

In analogy with (\ref{Ueqn}) we {\it define} a linear
combination $U$ of $\tilde u\equiv(u,\partial_i u)$ to be a characteristic
variable of the second order system with speed $\lambda$ in the
direction $n_i$ if it obeys
\begin{equation}
\label{2char}
\partial_t U=\lambda \partial_n U
+ \hbox{transversal derivatives}.
\end{equation}

We illustrate this again for the scalar field model.
The second order characteristic variables are 
\begin{eqnarray}
\label{2ndchar}
U_\pm &\equiv& \Pi\pm \partial_n \phi, \\
U_A &\equiv& \partial_A \phi, 
\end{eqnarray}
and obey
\begin{eqnarray}
\partial_t U_\pm &=& \pm \partial_n U_\pm + \partial^A U_A, \\
\partial_t U_A &=& \partial_A {1\over 2}(U_++U_-).
\end{eqnarray}
Note that one can obtain the non-zero speed characteristic variables
(\ref{2ndchar}) of the second order PDE system from those of the
pseudo-differential system, (\ref{scalarfourchar}), because both are
eigenvectors of essentially the same matrix $P_n$. In the second-order
system we neglect the transverse derivatives to find $P_n$, while in
the pseudo-differential system $n_i$ is defined as
$\omega_i/|\omega|$, and so the transversal derivatives vanish by
definition.

The definition (\ref{2char}) of characteristic variables for a
second-order in space system has two consequences which are not
apparent in the simple scalar field example. The first is that a
transversal derivative $\partial_A u$ is by definition a zero-speed
characteristic variable because we can interchange derivatives:
\begin{equation}
\partial_t(\partial_A u)=\partial_A(\partial_t u).
\end{equation}
The second is that all characteristic variables of a second-order
system are unique only up to addition of transversal derivatives. This
non-uniqueness will later be used up to write $\epsilon$ and $F^n$ in
terms of characteristic variables.

%%%%%%%%%%%%%%%%%%%%%%%%%%%%%%%%%%%%%%%%%%%%%%%%%%%%%%%%%%%%%%%%%%%%%%%%

\subsection{Symmetric hyperbolicity}

%%%%%%%%%%%%%%%%%%%%%%%%%%%%%%%%%%%%%%%%%%%%%%%%%%%%%%%%%%%%%%%%%%%%%%%%

We would like to generalise the energy (\ref{defE}) to a system that
is second order in space. As in the first order system, we require an
energy density and flux that obey the conservation law
\begin{equation}
\partial_t\epsilon=\partial_i F^i,
\end{equation}
and can be expressed in characteristic variables.

We demonstrate the second-order energy and maximally dissipative
boundary conditions for the scalar field. The scalar field admits a
positive definite conserved energy that is unique up to an overall
scale, with energy density
\begin{equation}
\label{scalarE}
\epsilon=\Pi^2+(\partial_i\phi)^2
\end{equation}
and flux
\begin{equation}
\label{scalardEdt}
F^i=2{\Pi}\partial^i\phi.
\end{equation}
The energy can be written as
\begin{equation}
\epsilon={1\over 2}(U_+^2+U_-^2)+U_A U^A
\end{equation}
and the flux in the direction $n_i$ normal to the boundary can be
written as
\begin{equation}
F^n={1\over 2}\left(U_+^2-U_-^2\right).
\end{equation}
This means that
\begin{equation}
U_+=\kappa\,U_-+f
\end{equation}
with $|\kappa|\le 1$ is a maximally dissipative boundary condition for
the system (\ref{wave1},\ref{wave2}). As the characteristic variables of
the second-order system contain derivatives, this is not an algebraic
but a first-order differential boundary condition, namely
\begin{equation}
\label{scalarBC}
(1-\kappa)\Pi+(1+\kappa)\partial_n \phi=f.
\end{equation}
We see that $\kappa=-1$ corresponds to an (inhomogeneous) Dirichlet
boundary condition on $\Pi$ and hence on $\phi$, $\kappa=1$ to Neumann, and
$\kappa=0$ to specifying the ingoing radiation (Sommerfeld).

The conserved energy with $\epsilon$ given by (\ref{scalarE}) can be
used to bound the $L_2$ norms of $\Pi$ and $\partial_i\phi$, but not
$||\phi||$ itself. To obtain a bound on $||\phi||$ as well, we can
use the energy given by
\begin{equation}
\tilde \epsilon=\Pi^2+(\partial_i\phi)^2+\alpha^2\phi^2
\end{equation}
with $\alpha>0$. 
Clearly this gives us 
\begin{equation}
||\Pi||\le \tilde E^{1/2}, \quad ||\partial_i\phi||\le \tilde E^{1/2},
\quad ||\phi||\le \alpha^{-1} \tilde E^{1/2},
\end{equation}
but with the disadvantage that $\tilde E$ is not strictly
conserved.  Rather
\begin{equation}
{d\tilde E\over dt}=
\int_{\partial\Omega} 2\Pi\partial^i\phi\,dS_i
+2\alpha^2 \int_\Omega \Pi\phi\,dV \le 2\alpha \tilde E
\end{equation}
where we have assumed maximally dissipative boundary conditions to
make the boundary term non-positive. This is an inefficient estimate,
but $\alpha$ can be made arbitrarily small.  Non-principal terms
in the equations can be dealt with in a similar manner: they give rise
to volume terms in $dE/dt$ that can be estimated as a multiple of $E$
itself.  (Note that the issues of bounding $||\phi||$ and of dealing
with non-principal terms are dealt with identically in second-order
and first-order systems.)

%%%%%%%%%%%%%%%%%%%%%%%%%%%%%%%%%%%%%%%%%%%%%%%%%%%%%%%%%%%%%%%%%%%%%%%%

\section{Constraint-preserving boundary conditions}
\label{section:cpbc}

%%%%%%%%%%%%%%%%%%%%%%%%%%%%%%%%%%%%%%%%%%%%%%%%%%%%%%%%%%%%%%%%%%%%%%%%

Here we summarise a general method of dealing with a system of
evolution equations that are subject to differential constraints
(\cite{CalabreseCPBC} and references there). As the constraints are by
assumption compatible with the evolution equations, they form a closed
evolution system. We now assume that this constraint system and the
main system are symmetric hyperbolic. The following discussion applies
then equally to first-order and second-order systems.

Rather than defining a notation for the general case, we consider the
case where the non-zero speed characteristic variables come in pairs
with speeds $\pm\lambda$, with $\lambda>0$, plus a number of
zero-speed variables. (This is necessarily true if there is no
preferred direction in space.) Focus on a pair of characteristic
constraint variables $C_\pm$ which obey
\begin{equation}
\partial_t C_\pm = \pm \lambda \partial_n C_\pm + \dots,
\end{equation}
where $\dots$ stands for transversal derivatives and any non-principal
terms. As the constraints are compatible with the evolution equations,
there exists a pair of characteristic variables of the main system
$U_\pm$ with the same speeds, that is
\begin{equation}
\partial_t U_\pm = \pm \lambda \partial_n U_\pm + \dots,
\end{equation}
such that 
\begin{equation}
\label{CUrel}
C_\pm = \partial_n U_\pm +\dots.
\end{equation}
The constraint energy is initially zero and we want it to remain
zero. One way of enforcing that is to formally impose homogeneous maximally
dissipative boundary conditions on the constraint system, that is
\begin{equation}
\label{Cdiss}
C_+=\kappa C_-
\end{equation}
for $|\kappa|\le 1$. If we now define
\begin{equation}
X\equiv U_+-\kappa U_-, 
\end{equation}
the boundary condition on the constraint system (\ref{Cdiss}) is
equivalent to the boundary condition
\begin{equation}
\partial_n X=\dots
\end{equation}
on the main system, taking into account the relation (\ref{CUrel}).
We now use the evolution equations for $U_\pm$ to replace $\partial_n$
with $\partial_t$, and find an evolution equation for $X$ which
contains only transversal derivatives (derivatives parallel to the boundary):
\begin{equation}
\label{Xdot}
\partial_t X=\dots
\end{equation}
The trick is to interpret this as an evolution equation for an
auxiliary variable $X$ which is defined only on the boundary. (In
general, there will be more than one pair of constraints, and a system
of variables $X$.) The boundary condition
\begin{equation}
\label{cpbc}
U_+=\kappa U_-+X
\end{equation}
has the form of a maximally dissipative boundary condition on $U$, but
it is one properly speaking only if $X$ can be treated as given. This
is the case only if the boundary system (\ref{Xdot}) decouples from
the bulk system.

One can show that, with maximally dissipative boundary conditions, 
\begin{eqnarray}
{d\over dt}||X||_{\partial\Omega}&\le& K_1 ||f||_{\partial\Omega}
+ K_2 ||u||_{\partial\Omega}, \\
{d\over dt}||u||_\Omega&\le& K_3 ||X||_{\partial\Omega}.
\end{eqnarray}
If the boundary system decouples from the bulk system, $K_2=0$ and the
growth of the boundary solution and hence of the bulk solution is
bounded by $f$. But a bound on $||u||_\Omega$ does not imply one on
$||u||_{\partial \Omega}$, and so for $K_2\ne 0$ well-posedness cannot
be proved using these energy techniques. 

%%%%%%%%%%%%%%%%%%%%%%%%%%%%%%%%%%%%%%%%%%%%%%%%%%%%%%%%%%%%%%%%%%%%%%%%

\section{The Maxwell equations}
\label{section:maxwell}

%%%%%%%%%%%%%%%%%%%%%%%%%%%%%%%%%%%%%%%%%%%%%%%%%%%%%%%%%%%%%%%%%%%%%%%%

%%%%%%%%%%%%%%%%%%%%%%%%%%%%%%%%%%%%%%%%%%%%%%%%%%%%%%%%%%%%%%%%%%%%%%%%

\subsection{Field equations}

%%%%%%%%%%%%%%%%%%%%%%%%%%%%%%%%%%%%%%%%%%%%%%%%%%%%%%%%%%%%%%%%%%%%%%%%

The Maxwell (electromagnetism, EM) equations in the presence of a
charge density $\rho$ and current density $j_i$ can be written as
the evolution equations
\begin{eqnarray}
\label{EM1}
\partial_t A_i &=& - E_i - \partial_i \psi, \\
\label{EM2}
\partial_t E_i &=& - \partial_j \partial^j A_i 
+ \partial_i\partial^j A_j - 4\pi j_i,
\end{eqnarray}
subject to a constraint
\begin{equation}
C_E \equiv \partial^i E_i-4\pi\rho=0.
\end{equation}
Here $\psi$, $j_i$ and $\rho$ can be treated as given functions,
subject to the charge conservation law $\partial_t \rho+\partial^i
j_i=0$. All indices are moved with the flat Cartesian spatial metric
$\delta_{ij}$, and staggered double indices are summed over.

The EM system would be a wave equation for $A_i$ were it not for the
grad div term $\partial_i \partial^j A_j$. This term can be eliminated
by making the offending divergence a new variable. This property makes
the EM system a nice toy model for the ADM formulation of the
Einstein equations \cite{KnappWalkerBaumgarte}. We therefore introduce
$\Gamma\equiv\partial^iA_i$. This gives rise to the new constraint
\begin{equation}
C_\Gamma \equiv \Gamma - \partial^i A_i = 0.
\end{equation}
We obtain an evolution equation for $\Gamma$ by taking the divergence
of the evolution equation for $A_i$, and add to it $bC_E$ with a
free coefficient $b$. We also add $a\partial_i C_\Gamma$ to the evolution
equation for $E_i$ with a free coefficient $a$. The resulting
system is
\begin{eqnarray}
\partial_t A_i &=& - E_i - \partial_i \psi, \\
\partial_t E_i &=& - \partial_j \partial^j A_i 
+ a\partial_i \Gamma + (1-a) \partial_i\partial^j A_j - 4\pi j_i, 
\quad {} \\
\partial_t \Gamma &=&(b-1) \partial_i E^i - b 4\pi\rho 
-\partial_i\partial^i \psi.
\end{eqnarray}
Note that for $a=0$, $\Gamma$ decouples and we recover the original
system. For $a=1$, $A_i$ obeys a wave equation. The constraint system
is
\begin{eqnarray}
\partial_t C_E &=& a \partial_i\partial^i C_\Gamma, \\
\partial_t C_\Gamma &=& b C_E.
\end{eqnarray}
In the following, we concentrate on the principal part. In the EM
system, this means neglecting the terms containing $\rho$, $j_i$ and
$\psi$. (For the EM system the principal part approximation happens to
be exact: it corresponds to vacuum electrodynamics in the gauge
$\psi=0$.)

We now use the Maxwell equations to illustrate three methods of
investigating well-posedness: differential reduction to first order,
pseudo-differential reduction, and making the original second-order
system symmetric hyperbolic. This is only for illustration: our
preferred method is the last one. We have introduced the
parameters $a$ and $b$ to clarify the presentation, and because
precisely analogous parameters exist for the ADM formulation of the
Einstein equations. We shall see that a natural choice of these
parameters would be $a=b=1$.

%%%%%%%%%%%%%%%%%%%%%%%%%%%%%%%%%%%%%%%%%%%%%%%%%%%%%%%%%%%%%%%%%%%%%%%%

\subsection{Differential reduction to first order}

%%%%%%%%%%%%%%%%%%%%%%%%%%%%%%%%%%%%%%%%%%%%%%%%%%%%%%%%%%%%%%%%%%%%%%%%

%%%%%%%%%%%%%%%%%%%%%%%%%%%%%%%%%%%%%%%%%%%%%%%%%%%%%%%%%%%%%%%%%%%%%%%%

\subsubsection{Field equations}

%%%%%%%%%%%%%%%%%%%%%%%%%%%%%%%%%%%%%%%%%%%%%%%%%%%%%%%%%%%%%%%%%%%%%%%%

To reduce the EM main system to first order, we introduce the
auxiliary variables $d_{ij}\equiv \partial_i A_j$. Because $A_i$ does
not appear undifferentiated in the field equations, this definition
itself does not appear as a new constraint, but its integrability
condition does:
\begin{equation}
C_{ijk}\equiv \partial_i d_{jk}- \partial_j d_{ik} = 0.
\end{equation}
The only term in the second-order system that does not translate
unambiguously into the new variables is $\partial_i\partial^j A_j$,
which can be written either as $\partial_i {d_j}^j$ or $\partial^j
d_{ij}$. We therefore parameterise this choice with a parameter
$c$. From the point of view of the new system, this can be described
as adding the constraint $-c{C_{ij}}^j$ to the evolution equation for
$E_i$ with a free coefficient $c$. The resulting system is (now neglecting
non-principal terms as discussed above)
\begin{eqnarray}
\partial_t d_{ij} &=& - \partial_i E_j, \\
\partial_t E_i &=& - \partial^j d_{ji} 
+ a\partial_i \Gamma + (1-a-c) \partial_i {d_j}^j + c\partial^j d_{ij},
\qquad {}
\\
\partial_t \Gamma &=&(b-1) \partial_i E^i .
\end{eqnarray}

To reduce the constraint system to first order, we introduce the new
variables $C_i\equiv \partial_i C_\Gamma$, where now
$C_\Gamma=\Gamma-{d_j}^j$. As $C_\Gamma$ does not appear in
undifferentiated form, this definition itself does not appear as a
constraint, but it gives rise to the integrability condition
\begin{equation} 
C_{ij}\equiv \partial_i C_j-\partial_j C_i=0. 
\end{equation}
The first-order constraint system is
\begin{eqnarray}
\partial_t C_E &=&  a \partial^i C_i
+(1-c)\partial^i {C_{ij}}^j, \\
\partial_t C_i &=& b \partial_i C_E, \\
\partial_t C_{ijk}&=&0, \\
\partial_t C_{ij}&=&0.
\end{eqnarray}

%%%%%%%%%%%%%%%%%%%%%%%%%%%%%%%%%%%%%%%%%%%%%%%%%%%%%%%%%%%%%%%%%%%%%%%%

\subsubsection{Strong hyperbolicity}

%%%%%%%%%%%%%%%%%%%%%%%%%%%%%%%%%%%%%%%%%%%%%%%%%%%%%%%%%%%%%%%%%%%%%%%%

We write the first-order system as
\begin{equation}
\label{firstordersystem}
\partial_t u=P_n\partial_n u+ P^A \partial_n A u.
\end{equation}
To diagonalise $P_n$, we split $d_{ij}$ into the tensor components
$d_{nn}$, $d_{qq}$, $d_{An}$, $d_{nA}$ and $d_{AB}$ with respect to a
fixed direction $n_i$, where the index $n$ means a contraction with
$n^i$, the index pair $qq$ means a contraction with $q^{ij}$, and
indices $A$, $B$, etc are the same as $i$, $j$ etc, but denote a
tensor that is transversal {\it and tracefree} on any index
pair. (Because it is transversal, it is tracefree with respect to both
$\delta_{ij}$ and $q_{ij}$.) Note that ${d_i}^i=d_{nn}+d_{qq}$. A
similar tensor decomposition is used for the other variables and
throughout this paper. We shall loosely refer to these objects as
scalars if they have no free (transversal) index, vectors if they have
one, and tensors if they have more than one.

By applying the same tensor decomposition to the evolution equations,
we find $P_n$ in this basis: it is block-diagonal with blocks
corresponding to the scalars $(d_{nn},d_{qq},E_n,\Gamma)$, transverse
vectors $(d_{nA},d_{An},E_A)$ and transverse tracefree tensors
$(d_{AB})$. The free transversal indices behave trivially, so that the
blocks we still have to diagonalise are quite small. 
The scalar block of $P_n$ is given by
\begin{eqnarray}
P_n d_{nn}&=& -E_n, \\
P_n d_{qq}&=& 0, \\
P_n E_n &=& a (\Gamma-d_{nn}-d_{qq}) + (1-c) d_{qq}, \\
P_n \Gamma &=& (b-1) E_n,
\end{eqnarray}
the vector block by
\begin{eqnarray}
P_n d_{nA}&=&-E_A, \\
P_n d_{An}&=&0, \\
\label{PnEA}
P_n E_A &=&-d_{nA}+cd_{An},
\end{eqnarray}
and the tensor block by $P_n d_{AB}=0$. The characteristic variables
obey $P_n U=U\Lambda$ and comprise the four scalars
\begin{eqnarray}
&& d_{qq}, \\
U_0 &\equiv& \Gamma + (b-1) d_{nn},\\
\label{Upmdef}
U_\pm &\equiv& a(\Gamma-d_{nn}-d_{qq})+(1-c)d_{qq}\pm \sqrt{ab} E_n, 
\end{eqnarray}
with speeds $(0,0,\pm \sqrt{ab})$, 
six vector components
\begin{eqnarray}
&& d_{An} ,\\
\label{UpmAdef}
U_{\pm A} &\equiv& d_{nA}-cd_{An}\mp E_A  ,
\end{eqnarray}
with speeds $(0,\pm 1)$, 
and three tensor components $d_{AB}$ with zero speed.
The characteristic variables for the constraints are
\begin{eqnarray}
&& C_A, \quad C_{ij}, \quad C_{ijk}, \label{Cpmdef} \\
C_\pm &\equiv& 
a C_n + (1-c)C_{ni}{}^i \pm\sqrt{ab}C_E,
\end{eqnarray}
with speeds $(0,\pm \sqrt{ab})$.  (We have not bothered to decompose
$C_{ij}$ and $C_{ijk}$ as all their components have zero speed.)  Both
the main and constraint system are strongly hyperbolic for
$ab>0$. Note that the {\it coefficients} of $\Gamma$ in $U_0$ etc. are
themselves left (row) eigenvectors of $P_n$.) The non-zero speed
variables contain terms which are themselves zero speed variables,
(for example, $U_\pm$ contains $d_{qq}$). 

In the base of characteristic variables the EM system is
\begin{eqnarray}
\label{first}
\partial_t d_{qq}&=&{1\over 2}\partial^A(U_{+A}-U_{-A}), \\
\partial_t U_0 &=& -{1\over 2}(b-1)\partial^A(U_{+A}-U_{-A}), \\
\nonumber\partial_t U_\pm &=&\pm \sqrt{ab} \, \partial_n U_\pm
-{ab+c-1\over 2}\partial^A(U_{+A}-U_{-A})  \\
&& \pm \sqrt{ab} \, \partial^A
\Bigl[{c\over 2}(U_{+A} \nonumber \\
&&+U_{-A})+(c^2-1)d_{An}\Bigr] \\
\partial_t U_{\pm A} &=& \pm \partial_n U_{\pm A}
+{c\over 2\sqrt{ab}}\partial_A (U_+-U_-)\nonumber \\
&&\pm \partial^B(d_{BA}-c\,d_{AB}) \nonumber \\ 
&&\mp \partial_A\Bigl[{ab+c-1\over2ab}(U_++U_-)+{1-c\over b}U_0
 \nonumber \\
&&+ \frac{c-1}{2ab}(2c-2+2a+ab)d_{qq}\Bigr], \\
\label{dAndot}
\partial_t d_{An} &=& -{1\over 2\sqrt{ab}}\partial_A (U_+-U_-), \\
\partial_t d_{AB} &=& {1\over 2}\partial_A
(U_{+B}-U_{-B}) \nonumber \\
&&-{1\over 4}q_{AB}\partial^C(U_{+C}-U_{-C}). 
\label{last}
\end{eqnarray}
Similarly, the constraint system in characteristic variables is
\begin{eqnarray}
\partial_t C_\pm &=& \pm \sqrt{ab} \,\partial_n C_\pm
\nonumber \\
&&\pm\sqrt{ab} \,\partial^A \left[aC_A+(1-c){C_{Aj}}^j\right], 
\qquad {} \\
\partial_t (aC_A) &=& {\sqrt{ab}\over 2}\partial_A(C_+-C_-), \\
\partial_t C_{ij}&=&0, \\
\partial_t C_{ijk}&=&0.
\end{eqnarray}

%%%%%%%%%%%%%%%%%%%%%%%%%%%%%%%%%%%%%%%%%%%%%%%%%%%%%%%%%%%%%%%%%%%%%%%%

\subsubsection{Symmetric hyperbolicity}

%%%%%%%%%%%%%%%%%%%%%%%%%%%%%%%%%%%%%%%%%%%%%%%%%%%%%%%%%%%%%%%%%%%%%%%%

The first-order EM system admits the conserved energy
\begin{equation}
\label{covariant_epsilon}
\epsilon = c_0 \epsilon_0 + c_1 \epsilon_1 ,
\end{equation}
where $c_0$ and $c_1$ are two arbitrary coefficients and 
\begin{eqnarray}
\epsilon_0 &=& E^i E_i + d^{ij}d_{ij} - c\, d^{ij}d_{ji}
- 2a\Gamma d_i{}^i \nonumber \\
&& +(c-1-ab+2a)(d_i{}^i)^2 , \\
\epsilon_1 &=& \left[\Gamma+(b-1)d_i{}^i\right]^2 .
\end{eqnarray}
The flux is
\begin{equation}
F^i = 2c_0 \left[ a(\Gamma-d_j{}^j) E^i +(1-c) d_j{}^j E^i
-(d^{ij}-c d^{ji})E_j\right]
\end{equation}
Note that $\epsilon_1$ has zero flux.
$\epsilon$ and $F^n$ can be written
in terms of characteristic variables as
\begin{eqnarray}
\nonumber \epsilon&=&c_0\left[d_{AB}d^{AB} -c d_{AB} d^{BA}
+(1-c^2) d_{An} d^{An} \right. \\
\nonumber&& \quad +{1\over 2} (U_{+A}U_+^A+U_{-A}U_-^A)
+{1\over 2ab}(U_+^2+U_-^2)
\nonumber \\
&& \quad \left. +\frac{c-1}{2ab}(4ad_{qq}U_0+(2-4a+ab-2c)d_{qq}{}^2)
\right] \nonumber \\
&& + \left(c_1-{a\over b}c_0\right)[U_0+(b-1)d_{qq}]^2 \\
\label{boundEM}
F^n &=& {1\over 2}c_0 \Bigl[(U_{+A}U_+^A-U_{-A}U_-^A)\nonumber \\
&&+{1\over\sqrt{ab}}(U_+^2-U_-^2)\Bigr].
\end{eqnarray}

Now we impose that $\epsilon$ is positive definite. We can write
\begin{equation}
\label{sincostrick}
d_{ij}d^{ij}-c \,
d_{ij}d^{ji}=(\cos\varphi d_{ij}-\sin\varphi d_{ji})^2
\end{equation}
where $\sin 2\varphi\equiv c$. This is positive definite for $|c|<
1$. Overall there are four positivity conditions:
\begin{eqnarray}
c_0&>&0, \quad c_1>\frac{3a^2}{3ab+2c-2}c_0, \nonumber \\
|c|&<& 1, \quad ab>\frac{2}{3}(1-c).
\label{EMineqs}
\end{eqnarray}

The covariant constraint energy density and flux, after fixing an
overall factor, are
\begin{eqnarray}
\epsilon_c &=& [aC_i+(1-c)C_{ij}{}^j]^2+ab C_E^2 
\nonumber \\
&&+w_1(C_{ij})^2+w_2(C_{ijk})^2, \\
F^i_c&=&2abC_E[aC_i+(1-c)C_{ij}{}^j].
\end{eqnarray}
It is unconditionally positive definite.
In terms of characteristic variables, 
\begin{eqnarray}
\epsilon_c &=& {1\over 2}(C_+^2+C_-^2) + [a C_A+(1-c)C_{Aj}{}^j]^2
\nonumber \\
&&+w_1(C_{ij})^2+w_2(C_{ijk})^2, \\
\label{boundEMc}
F^n_c&=& {1\over 2} \sqrt{ab} \left(C_+^2 - C_-^2\right).
\end{eqnarray}
The terms multiplied by $w_1$ and $w_2$ are strictly constant, and so
$w_1>0$ and $w_2>0$ are arbitrary.

For the main system and the constraint system to be symmetric
hyperbolic, the parameters of the system must obey the 
positivity conditions (\ref{EMineqs}). The strong hyperbolicity
condition $ab>0$, which must also be obeyed, is implied by these.

%%%%%%%%%%%%%%%%%%%%%%%%%%%%%%%%%%%%%%%%%%%%%%%%%%%%%%%%%%%%%%%%%%%%%%%%

\subsection{Pseudo-differential reduction to first order}

%%%%%%%%%%%%%%%%%%%%%%%%%%%%%%%%%%%%%%%%%%%%%%%%%%%%%%%%%%%%%%%%%%%%%%%%

%%%%%%%%%%%%%%%%%%%%%%%%%%%%%%%%%%%%%%%%%%%%%%%%%%%%%%%%%%%%%%%%%%%%%%%%

\subsubsection{Field equations}

%%%%%%%%%%%%%%%%%%%%%%%%%%%%%%%%%%%%%%%%%%%%%%%%%%%%%%%%%%%%%%%%%%%%%%%%

We now carry out the pseudo-differential reduction for
comparison. Defining $\hat u$ as the Fourier transform of $u$ in
space, and substituting this into the field equations we obtain
\begin{eqnarray}
\partial_t \hat A_i &=& - \hat E_i, \\
\partial_t \hat E_i &=&  \omega_j \omega^j \hat A_i 
+ ai\omega_i \hat \Gamma - (1-a) \omega_i\omega^j \hat A_j, \\
\partial_t \hat \Gamma &=&(b-1) i\omega_i \hat E^i .
\end{eqnarray}
The fact that the system is second order in space is indicated by the
fact that the highest power of $|\omega|$ on the right-hand side is
$|\omega^2|$. On the other hand, there is no power of $|\omega|$ on
the right-hand side of $\partial_t \hat A_i$. We
define new variables by absorbing powers of $|\omega|$ into some of
the Fourier transforms, namely
\begin{equation}
\hat a_i\equiv i|\omega|\hat A_i.
\end{equation}
In these variables, the ODE system in the Fourier variables becomes. 
\begin{eqnarray}
\partial_t \hat a_i &=& -i|\omega|\hat E_i, \\
\partial_t \hat E_i &=&  
i|\omega|\left[-\hat a_i 
+ a n_i \hat \Gamma 
+ (1-a)  n_i n^j \hat a_j\right], \\
\partial_t \hat \Gamma &=&i|\omega|(b-1)  n_i \hat E^i, 
\end{eqnarray}
where again $n_i \equiv \omega_i/|\omega|$. The right-hand sides are
now the symbols of first-order pseudo-differential operators, but only
the third is also the symbol of a differential operator.

%%%%%%%%%%%%%%%%%%%%%%%%%%%%%%%%%%%%%%%%%%%%%%%%%%%%%%%%%%%%%%%%%%%%%%%%

\subsubsection{Strong hyperbolicity}

%%%%%%%%%%%%%%%%%%%%%%%%%%%%%%%%%%%%%%%%%%%%%%%%%%%%%%%%%%%%%%%%%%%%%%%%

We can now find the characteristic variables of the
pseudo-differential system using the same tensor decomposition as
above. We write the pseudo-differential system as
\begin{equation}
\label{pseudodiffsystem}
\partial_t \hat u=i|\omega|P_n\hat u.
\end{equation}
$P_n$ in (\ref{pseudodiffsystem}) is a submatrix of $P_n$ in
(\ref{firstordersystem}), with $E_i\to \hat E_i$, $\Gamma\to\hat
\Gamma$, $\partial_n A_i\to i|\omega|\hat A_i\equiv \hat a_i$, but
$\partial_A A_i\to 0$ because there are no transversal derivatives in
the pseudo-differential framework. Therefore the rows and columns
corresponding to $d_{Ai}$ and $d_{qq}$ disappear. Note that with
these deletions, $c$ disappears from $P_n$: this must happen as $c$
parameterises an ambiguity of the differential reduction. The
characteristic variables are
\begin{eqnarray}
\hat U_0 &\equiv& \hat \Gamma + (b-1)\hat a_n, \\
\hat U_\pm &\equiv& a(\hat\Gamma-\hat a)\pm \sqrt{ab} \hat E_n , \\
\hat U_{\pm B} &\equiv& \hat a_B\mp \hat E_B , 
\end{eqnarray}
with speeds $(0,\pm\sqrt{ab},\pm 1)$. 

The Fourier transforms of the constraints are
\begin{equation}
\hat C_E=i|\omega|\hat E_n, 
\qquad \hat C_\Gamma=\hat \Gamma-\hat a_n.
\end{equation}
With
\begin{equation}
\hat c_\Gamma\equiv i|\omega|\hat C_\Gamma
\end{equation}
we have
\begin{eqnarray}
\partial_t \hat C_E&=&i|\omega|a\hat c_\Gamma, \\
\partial_t \hat c_\Gamma&=&i|\omega|b\hat C_E.
\end{eqnarray}
The characteristic variables are
\begin{equation}
\hat C_\pm\equiv a \hat c_\Gamma\pm \sqrt{ab}\hat C_E
= i|\omega|\hat U_\pm
\end{equation}
with speeds $\pm \sqrt{ab}$.
The pseudo-differential system is strongly hyperbolic for $ab>0$.

Because of the relation of the two matrices $P_n$ discussed above,
the characteristic variables for the pseudo-differential
reduction could be obtained from those for the differential reduction
by the replacements $d_{Ai}\to 0$, $d_{ni}\to \hat a_i$, $E_i\to \hat
E_i$, $\Gamma \to \hat \Gamma$, and $C_E \to \hat C_E$, $C_A\to 0$,
$C_n\to \hat C_\Gamma$, $C_{ij}\to 0$, $C_{ijk}\to 0$. 

%%%%%%%%%%%%%%%%%%%%%%%%%%%%%%%%%%%%%%%%%%%%%%%%%%%%%%%%%%%%%%%%%%%%%%%%

\subsection{Second-order in space, first order in time}

%%%%%%%%%%%%%%%%%%%%%%%%%%%%%%%%%%%%%%%%%%%%%%%%%%%%%%%%%%%%%%%%%%%%%%%%

%%%%%%%%%%%%%%%%%%%%%%%%%%%%%%%%%%%%%%%%%%%%%%%%%%%%%%%%%%%%%%%%%%%%%%%%

\subsubsection{Strong hyperbolicity}

%%%%%%%%%%%%%%%%%%%%%%%%%%%%%%%%%%%%%%%%%%%%%%%%%%%%%%%%%%%%%%%%%%%%%%%%

To construct the second-order characteristic variables, we consider
the principal part of the evolution of $\tilde u\equiv
(E_i,\partial_iA_j,\Gamma)$, which obey
\begin{equation}
\partial_t\tilde u=P_n\partial_n \tilde u + P^A \partial_A \tilde u.
\end{equation}
One might think that $P_n$ is the same matrix as in the differential
first order reduction. But consider, for example, Eq.~(\ref{PnEA}),
which is shorthand for
\begin{equation}
\partial_tE_A=-\partial_nd_{nA}+c\partial_nd_{An}+\hbox{transversal
derivatives}
\end{equation}
In the second-order system this becomes
\begin{equation}
\partial_tE_A=-\partial_n^2 A_A+c\partial_n\partial_A A_n+\hbox{transversal
derivatives},
\end{equation}
but we can commute derivatives and so lump $\partial_n\partial_A
A_n=\partial_A\partial_n A_n$ with the transversal derivatives. As
$\partial_t\partial_A A_n=\partial_A\partial_t A_n$ can also be
considered purely transversal, we can set both the row and column
corresponding to $\partial_A A_n$ in $P_n$ equal to zero. This is true
for all $\partial_A A_i$, and $\partial A C_\Gamma$ in the constraint
system. Therefore we are effectively diagonalising smaller matrices
$P_n$ for the main and constraint systems. These are identical (up to
variable names) to those in the pseudo-differential reduction.

The characteristic variables for the second-order EM system obtained
by diagonalising this reduced $P_n$ are
\begin{eqnarray}
U_0&\equiv& \Gamma+(b-1)\partial_n A_n, \\
U'_\pm&\equiv& a(\Gamma-\partial_n A_n)\pm \sqrt{ab} E_n, \\
U'_{\pm B}&\equiv&\partial_n A_B \mp E_B 
\end{eqnarray}
with speeds $(0,\pm\sqrt{ab},\pm 1)$
for the main system, and
\begin{equation}
\label{defCpm}
C_\pm\equiv a \partial_n C_\Gamma\pm \sqrt{ab}C_E
\end{equation}
with speeds $\pm \sqrt{ab}$ for the constraint system. To complete the
set of characteristic variables, we add the transversal derivatives
\begin{equation}
\partial_A A_i, \quad \partial_A C_\Gamma,
\end{equation}
which are by definition zero-speed variables.

%%%%%%%%%%%%%%%%%%%%%%%%%%%%%%%%%%%%%%%%%%%%%%%%%%%%%%%%%%%%%%%%%%%%%%%%

\subsubsection{Symmetric hyperbolicity}

%%%%%%%%%%%%%%%%%%%%%%%%%%%%%%%%%%%%%%%%%%%%%%%%%%%%%%%%%%%%%%%%%%%%%%%%

We obtain a conserved energy for the second-order system by writing
down the most general scalar $\epsilon$ that is quadratic in $\Gamma$,
$E_i$ and $\partial_i A_j$ and adjusting the free coefficients to
obtain conservation. The result is
\begin{eqnarray}
\epsilon&=&c_0\epsilon_0+c_1\epsilon_1+c_2\epsilon_2, \\
\epsilon_0&=&E_i^2+(\partial_i A_j)^2-2a\Gamma\partial^i A_i
\nonumber \\ &&
+(2a-1-ab)(\partial^i A_i)^2, \\
\epsilon_1&=&[\Gamma+(b-1)\partial^i A_i]^2, \\
\epsilon_2&=&(\partial^i A_i)^2-\partial_i A_j\partial^j A^i.
\end{eqnarray}
The flux is given by
\begin{eqnarray}
F_0^i&=&2\left\{[(a\gamma+(1-a)\partial^j A_j]E^i-\partial^iA^j
E_j\right\}, \\
F_2^i&=&2(\partial^i A^j E_j -\partial^j A_j E^i),
\end{eqnarray}
while $\epsilon_1$ has no flux. 
The constraint energy is unique up to an overall factor, and is given by
\begin{eqnarray}
\epsilon_c&=&a^2\partial_i C_\Gamma \partial^i C_\Gamma
+abC_E^2, \\
\label{Ecbound}
F^n_c&=& 2a^2bC_\Gamma \partial^i C_E.
\end{eqnarray}

If we add zero-speed terms $\partial_A A_i$ to $U'_\pm$ and $U'_{\pm A}$
as follows,
\begin{eqnarray}
\label{defUBpm}
U_\pm&\equiv& a(\Gamma-\partial_n A_n-\partial^BA_B)\pm \sqrt{ab} E_n
\nonumber \\
&&+(1-c) \partial^BA_B, \\
\label{defUpm}
U_{\pm B}&\equiv&\partial_n A_B \mp E_B -c\partial_B A_n,
\end{eqnarray}
where $c\equiv c_2/c_0$, we
can write the main flux in the form (\ref{boundEM}), and with
(\ref{defCpm}) we can write the constraint flux in the form
(\ref{boundEMc}). Furthermore, the evolution equations expressed in
the second-order characteristic variables are given by
(\ref{first}-\ref{last}), with the obvious replacements. This works
because the same constant $c$ arises twice in different
ways: in the first-order system it parameterises an arbitrariness of
writing the second-order field equations in terms of first-order
variables, and this is passed on to the energy. In the second-order
system there is no such ambiguity in the field equations, but $c$
arises as the free coefficient $c_2/c_0$ in the energy. 

%%%%%%%%%%%%%%%%%%%%%%%%%%%%%%%%%%%%%%%%%%%%%%%%%%%%%%%%%%%%%%%%%%%%%%%%

\subsection{Constraint-preserving boundary conditions}

%%%%%%%%%%%%%%%%%%%%%%%%%%%%%%%%%%%%%%%%%%%%%%%%%%%%%%%%%%%%%%%%%%%%%%%%

The following discussion applies equally to the first-order and
second-order symmetric hyperbolic systems. The constraint-preserving
boundary conditions we are about to derive contain only time
derivatives when applied to the first-order system, but both space and
time derivatives when applied to the second-order system. We shall
explicitly write down expressions for the first-order system. The
corresponding expressions for the second order system are obtained
by the replacement $d_{ij}\to \partial_i A_j$, and
interpreting the characteristic variables as those of the second-order
system. 

The maximally dissipative boundary conditions on the main variables
are
\begin{eqnarray}
\label{junk}
U_+-\kappa_1U_-&=&f, \\
\label{physicaldissEM}
U_{+A}-\kappa_2U_{-A}&=&f_A, 
\end{eqnarray}
where $|\kappa_1|\le 1$ and $|\kappa_2|\le 1$, and $f$ and $f_A$ are
given functions. We are only interested in solutions with vanishing
constraints. Therefore we formally impose homogeneous maximally
dissipative boundary conditions on the constraint system, that is
\begin{equation}
\label{dissconstrEM}
C_+-\kappa_3C_-=0
\end{equation}
with $|\kappa_3|\le 1$. We now show how these boundary conditions can
be made consistent, following the general procedure set out in
Sec.~\ref{section:cpbc}.

$C_\pm$ can be expressed in terms of the main characteristic variables
as
\begin{eqnarray}
C_{\pm} &=&
\partial_n U_\pm + \frac{1}{2}\partial^A
\Bigl[\mp\sqrt{ab}(U_{+A}-U_{-A})  \nonumber \\
&&+(c-1)(U_{+A}+U_{-A}+2c\, d_{An})\Bigr].
\label{Upmt}
\end{eqnarray}
We use the evolution equation for $U_\pm$ to replace $\partial_nU_\pm$
by $\partial_t U_\pm$. Eq.~(\ref{dissconstrEM}) then
becomes
\begin{eqnarray}
\label{XdotEM}
&&\partial_t (U_++\kappa_3U_-) \nonumber \\ &=&
\frac{(1+\kappa_3)(1-c)}{2}\partial^A(U_{+A}-U_{-A}) \\
&&+\frac{(1-\kappa_3)\sqrt{ab}}{2}\partial^A[U_{+A}+U_{-A}+2(c-1)d_{An}] .
\nonumber 
\end{eqnarray}
We impose this as an evolution equation for an auxiliary field
$X\equiv U_++\kappa_3U_-$ on the boundary. We then impose the boundary
condition (\ref{junk}) on $U_+$ with $\kappa_1=-\kappa_3$ and
$f=X$. We also impose the maximally dissipative boundary conditions
(\ref{physicaldissEM}) on $U_{+A}$, which represent the two
polarisations of ingoing radiation. $\kappa_2$ and $f_A$ must
therefore be chosen on physical grounds: stability only dictates
$|\kappa_2|\le 1$.

%%%%%%%%%%%%%%%%%%%%%%%%%%%%%%%%%%%%%%%%%%%%%%%%%%%%%%%%%%%%%%%%%%%%%%%%%%%

\subsubsection{Dirichlet boundary conditions}

%%%%%%%%%%%%%%%%%%%%%%%%%%%%%%%%%%%%%%%%%%%%%%%%%%%%%%%%%%%%%%%%%%%%%%%%%%%

As discussed in general in Sec.~\ref{section:cpbc}, one can
demonstrate well-posedness with these constraint-preserving boundary
conditions if the bulk system does not couple back to the main
system. There are two cases in which this can be achieved: For
$\kappa_3=1$, $d_{An}$ is eliminated from the boundary system, while
for $\kappa_3=-1$, its evolution equation(\ref{dAndot}) becomes part
of an expanded closed boundary system. For
$\kappa_3=\kappa_2=-\kappa_1=1$, the boundary system and boundary
conditions are
\begin{eqnarray}
\partial_t X&=&(1-c)\partial^A f_A, \\
U_++U_+&=&X, \\
U_{+A}-U_{-A}&=&f_A,
\end{eqnarray}
where $f_A$ is free data. 
The boundary condition on $U_\pm$ corresponds to Dirichlet
boundary conditions on $C_E$. 

%%%%%%%%%%%%%%%%%%%%%%%%%%%%%%%%%%%%%%%%%%%%%%%%%%%%%%%%%%%%%%%%%%%%%%%%%%%

\subsubsection{Neumann boundary conditions}

%%%%%%%%%%%%%%%%%%%%%%%%%%%%%%%%%%%%%%%%%%%%%%%%%%%%%%%%%%%%%%%%%%%%%%%%%%%

For $\kappa_3=\kappa_2=-\kappa_1=-1$, the boundary system and the boundary conditions
for the main system are
\begin{eqnarray}
\partial_t X&=&\sqrt{ab}\partial^A [f_A+2(c-1)d_{An}], \\
\partial_t d_{An}&=&-{1\over 2\sqrt{ab}}\partial_A X, \\
U_+-U_-&=&X, \\
\label{UANeumann}
U_{+A}+U_{-A}&=&f_A.
\end{eqnarray}
The boundary condition on $U_\pm$ corresponds to Neumann boundary
conditions on $C_\Gamma$. 

The boundary system is symmetric hyperbolic with the energy density,
flux, and source term
\begin{eqnarray}
\epsilon_b&=&X^2+ 4ab(1-c) (d_{An})^2, \\
\partial_t\epsilon_b&=&\partial_A F^A_b+s_b, \\
F^A_b&=& 4\sqrt{ab}(c-1)Xd_{An}, \\
s_b&=& 2\sqrt{ab}X\partial^A f_A.
\end{eqnarray}
The characteristic variables of the boundary system in the
direction $m_A$ are
\begin{equation}
\label{Xpm}
X_{\pm}\equiv X \mp 2\sqrt{ab(1-c)} \,d_{mn},
\end{equation}
with speeds $\pm\sqrt{1-c}$. The boundary energy is controlled by
\begin{eqnarray}
\label{EMboundaryEdot}
{dE_b\over dt}&=&\int_{\partial\Omega}s\, dS
+\int_{\partial\partial\Omega}F^m_b\, ds, \\
\label{EMboundaryflux}
F^m_b &=&{1\over 2}\sqrt{1-c}(X_+^2-X_-^2),
\end{eqnarray}
where $m_A$ is the normal to $\partial\partial\Omega$ in
$\partial\Omega$.

For a smooth boundary ($\partial\partial\Omega=0$), the boundary
energy can therefore be estimated by
\begin{equation}
\label{EMboundaryestimate}
{dE_b\over dt} \le 2\sqrt{ab}\,||X||\, ||\partial^Af_A||
\le 2\sqrt{ab}\,\sqrt{E_b}\, ||\partial^Af_A||
\end{equation}
For a cubic boundary, one also has to consider the
$\partial\partial\Omega$ terms at the edges of each face. From
(\ref{UANeumann}) and (\ref{UpmAdef}) we have
\begin{equation}
\label{dmndata}
d_{mn}={f^{(m)}_n+cf^{(n)}_m \over 1-c^2}
\end{equation}
where $f^{(n)}_A$ are the free data on the boundary with normal
$n_i$. They are already given. This leaves us no choice but to impose
Dirichlet boundary conditions given by (\ref{dmndata}) on $d_{mn}$ at
the edge with normal $m_A$ of the face with normal $n_i$. From
(\ref{Xpm}) and (\ref{EMboundaryflux}) we see that these are maximally
dissipative boundary conditions on the system on each face, so that
the estimate (\ref{EMboundaryestimate}) still holds. Therefore the
coupled bulk/boundary system is well-posed in the Neumann case.

%%%%%%%%%%%%%%%%%%%%%%%%%%%%%%%%%%%%%%%%%%%%%%%%%%%%%%%%%%%%%%%%%%%%%%%%%%%

\subsubsection{Laplace-Fourier analysis}

%%%%%%%%%%%%%%%%%%%%%%%%%%%%%%%%%%%%%%%%%%%%%%%%%%%%%%%%%%%%%%%%%%%%%%%%%%%

With our current methods we can decouple the boundary system
from the bulk system, and hence prove well-posedness of the
initial-boundary value problem,
only for the two extreme values $\kappa_3=\pm 1$. The same was
found previously in imposing constraint-preserving boundary conditions
on the linearised Einstein-Christoffel system \cite{CalabreseCPBC} and
on the full Einstein equations in harmonic gauge
\cite{SzilagyiWinicour}. 
But the lack of a proof does not rule out that for example the
Sommerfeld boundary conditions with $\kappa_3=0$ could also be well-posed.

To investigate this possibility further, we now use the
Laplace-Fourier method to explicitly check for modes that grow
arbitrarily rapidly. The EM equations are untypical in that they are
linear with constant coefficients. For a quasilinear system such as
the Einstein equations, we use the high frequency, or frozen
coefficients, approximation, and consider only the principal part of
the evolution equations and constraints, thus recovering a linear
system with constant coefficients. In the frozen coefficients
approximation we assume that the linearised perturbation varies over
space and time scales much smaller than those given by 
the background solution and
the numerical domain. For consistency we must therefore assume that
the domain is a half space and that the boundary is a plane. Without
loss of generality, we assume that the evolution domain is
$-\infty<x^1\le 0$ and $-\infty<x^A\equiv(x^2,x^3)<\infty$. The
general solution with homogeneous boundary conditions can then be
written as a sum of modes of the form
\begin{equation}
u(x^i,t)=e^{st+i\omega_Ax^A+\lambda x^1}v
\end{equation}
where $v$ is a constant eigenvector to be determined. (If the
corresponding eigenvalue $\lambda$ was degenerate, $v$ would be a
polynomial in $x^1$ rather than constant, but this does not happen in
for the EM system.)  If the boundary conditions are inhomogeneous,
we must add to this ansatz a particular integral that does not concern us
here because its growth is controlled by the boundary data. We are
interested only in modes that are square-integrable over space at any
moment in time. Therefore we assume that $\omega_A$ is real, and that
${\rm Re}\lambda\ge 0$. $s$ and $\lambda$ will in general be complex.

If a mode of this form exists for some $(s,\omega_A,\lambda,v)$, then
one exists also for $(ks,k\omega_A,k\lambda,v)$ for any
$k>0$. Therefore, if any growing mode exists, there are growing modes
with arbitrarily large growth rates $ks$, and the problem is
ill-posed. A necessary condition for well-posedness of the
initial-boundary value problem is therefore that the homogeneous
boundary conditions rule out the existence of any mode with ${\rm
Re}s>0$ for all real $\omega_A$ and all $\lambda$ with ${\rm
Re}\lambda\ge 0$.

We now examine our proposed constraint-preserving boundary conditions
for the EM system. To simplify the algebra, we restrict ourselves to
the case $a=b=1$, $c=0$ (which is symmetric hyperbolic). The evolution
equations for the characteristic variables simplify to
\begin{eqnarray}
\partial_t d_{qq}&=&{1\over 2}\partial^A(U_{+A}-U_{-A}), \\
\partial_t U_0 &=&0, \\
\partial_t d_{An} &=& -{1\over 2}\partial_A (U_+-U_-), \\
\partial_t d_{AB} &=& {1\over 2}\partial_A
(U_{+B}-U_{-B})-{1\over 4}q_{AB}\partial^C(U_{+C}-U_{-C}), \\
\partial_t U_\pm &=&\pm \partial_n U_\pm
\mp\partial^Ad_{An}, \\
\partial_t U_{\pm A} &=& \pm \partial_n U_{\pm A}
\pm \partial^Bd_{BA}
\mp \partial_A U_0 \pm \frac{1}{2}\partial^A d_{qq} .
\end{eqnarray}
(Once again, we could trivially rewrite this for the second-order system).
The constraint-preserving boundary conditions are
\begin{eqnarray}
\label{bound1}
\partial_t(U_++\kappa_3U_-)&=&\partial^A\Bigl[
U_{+A}-\kappa_3U_{-A} \nonumber \\ &&
-(1-\kappa_3)d_{An}\Bigr], \\
\label{bound2}U_{+A}-\kappa_2U_{-A}&=&0.
\end{eqnarray}
Using the Laplace-Fourier ansatz in the evolution equations, we obtain
a linear eigenvalue problem for $v$ with eigenvalue
$\lambda(s,|\omega|)$. Its solution is
\begin{eqnarray}
\left(\begin{array}{c}
U_+ \\ U_-
\end{array}\right)
&=&\gamma_1e^{st+i\omega_Ax^A+\lambda x^1}\left(\begin{array}{c}
1 \\ \psi^2
\end{array}\right),  \label{modeform1}\\ 
\left(\begin{array}{c}
U_{A+} \\ U_{A-}
\end{array}\right)
&=&\gamma_{2A}e^{st+i\omega_Ax^A+\lambda x^1}\left(\begin{array}{c}
1 \\ \psi^2
\end{array}\right),
\label{modeform2}
\end{eqnarray}
where $\lambda=\sqrt{s^2+|\omega|^2}$ and $\psi=\sqrt{z^2+1}-z$ with
$z\equiv s/|\omega|$. $\gamma_1$ and $\gamma_{2A}$ are free
coefficients. We ensure that ${\rm Re}\lambda>0$ for ${\rm
Re}s>0$ by putting the branch cut for the square root on the negative
real axis. 

Applying the boundary condition (\ref{bound2}) to (\ref{modeform2}) gives
the algebraic condition $\gamma_{2A}(\kappa_2\psi^2-1)=0$. This means that
$\gamma_{2A}=0$ unless $\kappa_2\psi^2(z)=1$. Now $\psi^2(z)$ maps the
right half-plane into the unit disk with the negative real axis
removed. Therefore, $\kappa_2\psi^2(z)=1$ for ${\rm Re}z>0$ 
is ruled out if $|\kappa_2|\le 1$. The boundary condition
(\ref{bound1}) applied to (\ref{modeform2})
gives a more complicated algebraic condition linking $\gamma_1$ and
$\omega^A\gamma_{2A}$.  However, with $\gamma_{2A}=0$ this simplifies,
after some algebra, to $\gamma_1(\kappa_3\psi^2-1)=0$, and so we must also
impose $|\kappa_3|\le1$.

Our calculation has shown that the initial-boundary value problem has
no growing square-integrable modes of the form
(\ref{modeform1},\ref{modeform2}) for $|\kappa_2|\le 1$ and
$|\kappa_3|\le 1$.  This is consistent with our energy method proof
that the EM initial-boundary value problem is well-posed for
$\kappa_3=\pm 1$ and $|\kappa_2|\le 1$, but it also suggests that the
problem is well-posed in the more general case, and we expect that a
proof of this can be found using a modified energy. 

%%%%%%%%%%%%%%%%%%%%%%%%%%%%%%%%%%%%%%%%%%%%%%%%%%%%%%%%%%%%%%%%%%%%%%%%

\section{Conclusions}

%%%%%%%%%%%%%%%%%%%%%%%%%%%%%%%%%%%%%%%%%%%%%%%%%%%%%%%%%%%%%%%%%%%%%%%%

We have proposed a new method for setting up a well-posed
initial-boundary value problem for evolution systems, such as variants
of the ADM equations, that are first order in time but second order in
space. One might call it the second-order energy method. It involves
two steps:

1. Find a complete set of characteristic variables for any given
direction that obey Eq. (\ref{2char}). They are not yet unique. This
is our definition of strong hyperbolicity.

2. Find a conserved covariant energy, and express it and the
corresponding flux in terms of the characteristic variables. This can
be done for a unique set of characteristic variables. This is our
definition of symmetric hyperbolicity.

Our definition of strong hyperbolicity for a second-order system is
equivalent to that using a pseudo-differential reduction to first order
\cite{KreissOrtiz,NOR}, as both definitions require the same matrix to
be diagonalisable. But the pseudo-differential method cannot be extended
to symmetric hyperbolicity, roughly speaking because Fourier
transforms do not allow for boundaries in space.  Our definition of
symmetric hyperbolicity for a second-order system is similar to that
for the differential reduction to first order, but it does not
introduce auxiliary variables and constraints, and so does not enlarge
the solution space. 

We have worked through both types of first-order reduction, as well as
the second-order energy method, in the example of electromagnetism, in
order to illustrate the rather subtle similarities and differences.
In a companion paper \cite{bssn2} we shall use the energy method to
prove symmetric hyperbolicity and suggest constraint-preserving
boundary conditions for variants of the ADM equations.

%%%%%%%%%%%%%%%%%%%%%%%%%%%%%%%%%%%%%%%%%%%%%%%%%%%%%%%%%%%%%%%%%%%%%%%%

\acknowledgments

The authors would like thank G Nagy, O Ortiz and O Reula for
communicating a draft paper and for discussions. CG would like to
thank the Kavli Institute for Theoretical Physics for hospitality
while this work was begun, and T Baumgarte, G Calabrese, H Friedrich,
L Kidder, L Lindblom, O Sarbach, M Scheel, D Shoemaker and J Vickers
for helpful discussions.
JMM was supported by the Comunidad Aut\'onoma de Madrid and Fondo
Social Europeo and also in part by the Spanish MCYT under the
research project BFM2002-04031-C02-02.

%%%%%%%%%%%%%%%%%%%%%%%%%%%%%%%%%%%%%%%%%%%%%%%%%%%%%%%%%%%%%%%%%%%%%%%%

\begin{appendix}

%%%%%%%%%%%%%%%%%%%%%%%%%%%%%%%%%%%%%%%%%%%%%%%%%%%%%%%%%%%%%%%%%%%%%%%%

\section{An alternative first-order reduction}

Lindblom et al \cite{Lindblometal} have imposed
constraint-preserving boundary conditions of the Sommerfeld type on a
symmetric hyperbolic first-order reduction of the EM system (without
proving well-posedness of the resulting initial-boundary value
problem). They consider the following first-order reduction of the EM
system:
\begin{eqnarray}
\partial_t E_i &=& \partial^j (d_{ij}-d_{ji})+{\gamma_1\over
2}(\partial_i {d_j}^j-\partial^jd_{ij}), \\
\partial_t d_{ij} &=& -\partial_i E_j+\gamma_2 \delta_{ij}\partial^k
E_k.
\end{eqnarray}
This corresponds to our first-order reduction with $a=0$ (and $\Gamma$
therefore not part of the system) and $c=1-(\gamma_1/2)$, plus
the additional term parameterised by $\gamma_2$. The latter plays a
similar role to our parameter $b$, in adding the constraint $C_E$ to
the evolution equation for ${d_i}^i(=\Gamma)$. They find that the
system is strongly hyperbolic for $\gamma_1\gamma_2>0$, and symmetric
hyperbolic for $0<\gamma_1<4$ and $\gamma_2>1/3$. However, this
first-order reduction has as its second-order counterpart only the
original EM system (\ref{EM1},\ref{EM2}), which is only weakly
hyperbolic. The term multiplied by $\gamma_1$ vanishes identically
because partial derivatives can be commuted, and the term $\gamma_2$
cannot arise if we evolve $A_i$ rather than $d_{ij}$. In hindsight,
one can see that both terms have second-order counterparts as long as
we have the trace of $d_{ij}$ as an auxiliary variable: the rest of
$d_{ij}$ is not required.

\end{appendix}

%%%%%%%%%%%%%%%%%%%%%%%%%%%%%%%%%%%%%%%%%%%%%%%%%%%%%%%%%%%%%%%%%%%%%%%%

%%%%%%%%%%%%%%%%%%%%%%%%%%%%%%%%%%%%%%%%%%%%%%%%%%%%%%%%%%%%%%%%%%%%%%%%


\begin{thebibliography}{}

\bibitem{KreissOrtiz} H Kreiss and OE Ortiz, in {\it The conformal
structure of spacetimes: Geometry, Analysis, Numerics}, eds J
Frauendiener and H Friedrich, Springer (Lecture Notes in Physics vol. 604),
Heidelberg, 2002.  

\bibitem{NOR} G Nagy, OE Ortiz and OA Reula, Strongly
hyperbolic second order Einstein's evolution equation, in preparation.

\bibitem{GustafssonKreiss} B Gustafsson, H-O Kreiss and J Oliger, {\it
Time-dependent problems and difference methods}, John Wiley, New York
1995.

\bibitem{Stewart} J. M. Stewart, Class. Quantum Grav. {\bf 15}, 2865
  (1998).

\bibitem{Reula} O Reula, Living Reviews in Relativity, {\bf 1998}-3.

\bibitem{CalabreseSarbach} G Calabrese and O Sarbach,
  J. Math. Phys. {\bf 44}, 3888 (2003). 

\bibitem{CalabreseCPBC} G Calabrese et al, Commun. Math. Phys. {\bf
240}, 377 (2003).

\bibitem{KnappWalkerBaumgarte} AM Knapp, EJ Walker and TW Baumgarte,
Phys. Rev. D {\bf 65}, 064031 (2002).

\bibitem{SzilagyiWinicour} B Szilagyi and J Winicour, Phys. Rev. D
{\bf 68}, 041501 (2003).

\bibitem{bssn2} C Gundlach and J M Mart\'\i n-Garc\'\i a, Symmetric
hyperbolic second-order Einstein equations, submitted to
Phys. Rev. D. 

\bibitem{Lindblometal} L Lindblom et al, Controlling the growth of
constraints in hyperbolic evolution systems, e-print gr-qc/0402027.

\end{thebibliography}
\end{document}